\documentstyle[aps,preprint,12pt]{revtex}
\tighten
\begin{document}
\title{Monte Carlo and Molecular Dynamics Simulation of the Glass Transition of
Polymers}
\author{Kurt Binder, J\"{o}rg Baschnagel, Christoph Bennemann and Wolfgang Paul}
\address{Institut f\"{u}r Physik, Johannes-Gutenberg-Universit\"{a}t, Mainz D-55099\\
Mainz, Staudinger Weg 7, Germany}
\maketitle

\begin{abstract}
Two coarse-grained models for polymer chains in dense glass-forming polymer
melts are studied by computer simulation: the bond fluctuation model on a
simple cubic lattice where a bond-length potential favors long bonds is
treated by dynamic Monte Carlo methods, and a bead-spring model in the
continuum with a Lennard-Jones potential between the beads is treated by
Molecular Dynamics. While the dynamics of both models differ for short
length scales and associated time scales, on mesoscopic spatial and temporal
scales both models behave similarly. In particular, the mode coupling theory
of the glass transition can be used to interpret the slowing down of the
undercooled polymer melt. For the off-lattice model, the approach to the
critical point of mode coupling is both studied for constant pressure and
for constant volume. The lattice model allows a test of the Gibbs-Di Marzio
entropy theory of the glass transition, and our finding is that although the
entropy does decrease significantly, there is no ``entropy catastrophe''
where the fluid entropy would turn negative. Finally, an outlook on
confinement effects on the glass transition in thin film geometry is given.

\vspace{5mm}

\noindent Contribution to the {\sf II Workshop on Non Equilibrium Phenomena
in Supercooled Fluids, Glasses and Amorphous Materials}, Pisa 1998, submitted
to {\em J. Phys.: Condens. Matter} on Sept.\ 27, 1998
\end{abstract}

\section{Introduction}

The glass transition from the supercooled melt to the amorphous solid still
is a puzzle \cite{1,2,3,4}. While the viscosity $\eta (T)$ increases about
15 orders of magnitude ($T_{g}$ is empirically defined by $\eta
(T=T_{g})=10^{13}$ Poise) in a narrow temperature interval, the change in
the static structure factor $S(q)\quad $($q$ is the scattering wavenumber)
is rather minor, see e.g. \cite{5}. What is then the physical basis that
many different systems show a nearly universal phenomenology of relaxation
\{e. g., increase of relaxation time $\tau $ or $\eta (T)\left[ \eta
(T)\propto \tau \right] $ described by the Vogel-Fulcher-Tammann (VFT) law, $%
\ln \tau \propto \ln \eta \propto E_{\text{VF}}/(T-T_{\text{VF}})$ \cite{1},
relaxation functions obey the Kohlrausch-Williams-Watts (KWW) law, $\varphi
(t)\propto \exp \left[ -(t/\tau )^{\beta }\right] $, etc. \cite{1}\}?

The answer to this question is controversial: e.g., there is the idea of an
``entropy catastrophe'' static phase transition underlying the glass
transition \{the configurational entropy of the undercooled fluid vanishes
at $T_{0}<T_{g}$\cite{6}, with possibly \cite{7} $T_{0}=T_{\text{VF}}$\}.
Mode coupling theory \cite{2}, on the other hand, yields a (rounded \cite{8}%
) dynamical transition from an ergodic fluid (atoms easily can escape from
the cages formed by their neighbors) at a ``critical'' temperature $%
T_{c}>T_{g}$ to a nonergodic state (``structural arrest''). Still other
concepts attribute the increase of $\tau $ upon cooling to the increase of a
glass correlation length \cite{9}, in analogy with spin glasses \cite{10}.

Polymer melts can be held in very good metastable equilibrium in their
supercooled state, and their $T_{g}$'s occur in an experimentally convenient
range, and thus a wealth of experimental data exists. Thus it is desirable
to approach these systems also by theoretical modeling via computer
simulation \cite{11}. However, due to the large size of these polymer coils
(they exhibit nontrivial structure from the \AA\ scale of covalent bonds up
to the gyration radius which can be 100 \AA) and the many decades of time
scales spanned by the relaxation times of their motion simulation of
polymers is very difficult, and requires to use simplified efficient models 
\cite{11}. Two such models will be defined and compared in the next section,
while Sec. 3 describes the tests of mode coupling theory performed with
these models. Sec. 4 briefly describes tests of other theories and contains
also some concluding remarks.

\section{Coarse-grained models for polymers on the lattice and in the
continuum}

The basic idea is to eliminate both the structural details on very small
scales and the associated very fast dynamics by using a coarse-graining
along the chemical sequence of the chains \cite{11}: $n\approx 3-5$
successive chemical units are mapped into one effective bond, connecting two
effective monomers. These effective bonds may be defined on a lattice as
well as in the continuum. In the lattice case, the effective monomers are
formed from all 8 sites of an elementary cube on the lattice, and no site
may belong to more than one monomer (excluded volume interaction). Bond
lengths in this ``bond fluctuation model'' \cite{11} are taken in the range
from 2 to $\sqrt{10}$ lattice spacings, and bond vectors $\vec{b}=(\pm
3,0,0) $ [or permutations thereof] are taken to be in the ground state,
while all other choices of the bond vector represent excited states, which
an energy cost $\varepsilon (=1;$ also k$_{\text{B}}\equiv 1$). This bond
length potential may be thought of as representing the original atomistic
intrachain potentials in a coarse-grained way. Since this particular choice
of the potential has the effect that each bond that is in the ground state
blocks the 4 sites between its adjoining effective monomers from further
occupation, it has the effect that free volume in the dense system is
``wasted'' and a ``geometric frustration'' is created - at high densities
not enough free volume is available that all bonds manage to get into their
ground state as $T\rightarrow 0$ \cite{12}. This model can be simulated very
efficiently, and for chain lengths N as small as $N=10$ properties typical
of polymers (e g. gyration radius $\left\langle R_{g}^{2}\right\rangle
^{1/2}\propto \sqrt{N}$) are reproduced. A density of $\phi =0.533$ of
occupied sites corresponds to a dense melt \cite{12,13}.

Dynamics is introduced by randomly selecting a monomer of a chain and by
attempting to move it by one lattice unit in a randomly chosen direction. Of
course, the move is only carried out if the new position satisfies excluded
volume and bond length constraints, and the possible energy change is
considered with the usual Metropolis criterion \cite{11,12}. It is thought
that this ``random hopping'' of effective monomers in a coarse-grained sense
corresponds to the random hops of chemical bonds over barriers in the
torsional potential, because by such moves pieces of a chain can rotate
relative to each other, and thus are responsible for the relaxation of chain
configurations \cite{11,14}. If one carries out such a mapping of a
realistic (i.e., atomistic) chain model on the bond fluctuation model
literally, one estimates that one attempted Monte Carlo step (MCS) per
monomer corresponds to 10$^{-13}$ sec, and the lattice unit to a distance of
2\AA\ \cite{14}. Typically, runs are carried out for systems of $L\times
L\times L$ with periodic boundary conditions and $L=30$, averaging over
16-160 independent ``replicas'' of the system over a time of up to 10$^{7}$
MCS. Thus even for the time window of this coarse-grained model we are
restricted to $t\approx 1\mu \sec$ (while molecular dynamics simulations
of atomistic models can deal with $t$%
\mbox{$<$}%
1nsec only). However, one important aspect is that one can equilibrate the
system configurations also with artificial moves, which have no counterpart
in the dynamics of real chains, such as the ``slithering snake'' algorithm 
\cite{11,14}, and thus gain several orders of magnitude in time due to
faster relaxation. In this way, configurations of the well equilibrated
melts are prepared at rather low temperatures, which serve as initial states
for runs with the ``random hopping'' algorithm studying the dynamics of
these melts.

Alternatively, we treat a continuum model where an effective bond is
represented by a (finitely extensible) spring, and the effective monomers
are the beads of this bead-spring model. There is a Lennard-Jones
interaction between any pair of beads cut off at a distance $r_{c}=2\times
2^{1/6}\sigma $ \cite{15},

\begin{equation}
U_{\text{LJ}}(r)=4\varepsilon \left[ (\sigma /r)^{12}-(\sigma /r)^{6}\right]
+c,  \label{eq1}
\end{equation}

the constant c being chosen such that U$_{\text{LJ}}(r_{c})=0$. Choosing
here units of length and temperature such that $\sigma =1$, $\varepsilon =1$%
, the springs are described by

\begin{equation}
U(r)=-\frac{1}{2}kR_{0}^{2}\ln \left[ 1-(r/R_{0})^{2}\right] ,\qquad
R_{0}=1.5,\quad k=30.  \label{eq2}
\end{equation}

While U$_{\text{LJ}}(r)$ has its preferred distance at a minimum for $r_{\min }
=2^{1/6}$, the minimum of the potential between neighboring monomers
along the chain (given by the sum of Eqs. (\ref{eq1}), (\ref{eq2})) occurs
for $r_{0}\approx 0.96$: although the bead-spring model is fully flexible,
this incompatibility of $r_{0}/r_{\min }$ with any standard crystal
structure prevents crystallization of our model, and thus creates the
``frustration'' in the fluid necessary for the formation of a glass upon
cooling.

This off-lattice model is simulated by NpT-molecular-dynamics, choosing also
a chain length of $N=10$, and altogether $M=1200$ monomers in the system 
\cite{15}. Note that the constant pressure ensemble \cite{16} is used for
equilibration only - for a study of dynamic properties a clock is set to
zero and runs are started at $T=$const (using a suitable Nos\'{e}-Hoover
thermostat \cite{17}) which produce a dynamical behavior practically
indistinguishable from the microcanonical ensemble \cite{15}.

Fig. 1 shows that one obtains the structure factor $S(q)$ of the amorphous
polymer melt in close similarity with the experiment \cite{5}. If one
studies the volume $v(T)$ per effective monomer in a slow cooling run,
starting at $p=1$ from a well-equilibrated configuration at $T=0.6$ and
lowering $T$ every 500 000 MD time steps by 0.02, one finds a rather
well-defined kink at $T_{g}\approx 0.41$, while a fit of the self-diffusion
constant to the Vogel--Fulcher law yields $T_{\text{VF}}\approx 0.33$ \cite
{18}. Thus the model does exhibit a glass transition, as expected.

The bond fluctuation model also yields $S(q)\approx 0$ for small $q$ due to
the very small compressibility of the polymer melt, and then a broad peak
similar to the ``amorphous halo'' of Fig.~1 occurs, at $q\approx 3$ (in
units of the lattice spacing for this model; physically this corresponds to
about 1.5 $\mbox{\AA}^{-1}$) \cite{13}. The peak position and shape also change
very little with temperature. However, the second shallow and
temperature-independent peak of Fig. 1 at still larger q, representing
intra-chain correlations, is less well reproduced \cite{13}. This is not
surprising, since $q=2\pi $ corresponds in real space to one lattice unit. A
further disadvantage of the lattice model is that one can study the glass
transition only at constant volume, while in the off-lattice model both
constant volume and constant pressure studies were performed \cite{18}.

\section{Testing the mode coupling theory (MCT) of the glass transition}

Fig. 2 shows the intermediate incoherent dynamic structure factor $F_{q}(t)$%
, defined as

\begin{equation}
F_{q}(t)=\frac{1}{M}\sum\limits_{i=1}^{M}\left\langle \exp \{i\vec{q}\cdot [%
\vec{r}_{i}(t)-\vec{r}_{i}(0)]\}\right\rangle ,  \label{eq3}
\end{equation}

where the sum is over all $M$ monomers in the system (which are at position $%
\vec{r}_{i}(t)$ at time $t$). Pronounced two-step relaxation is seen, the
second step (``$\alpha $-relaxation'') satisfies the
time-temperature-superposition principle, while the first step (``beta
relaxation'') does not \cite{15}. The relaxation time $\tau _{q}$ is both
compatible with the Vogel-Fulcher law with $T_{\text{VF}}\approx 0.33$, and
with a power law (Fig. 3) but with a q-dependent exponent $\gamma _{q}$ \cite
{18}. While the consistency with a power law is evidence in favor of MCT 
\cite{2}, the $q$-dependence of the exponent $\gamma _{q}$ is not. In
addition, one sees that the data deviate from the power law both for $%
T/T_{c}-1\gtrsim 0.4$ (as expected, since one has left the asymptotic
region) and for $T/T_{c}-1\lesssim 0.04$ (which can be attributed to
``hopping processes'' by which effective monomers can escape from the cage
formed by their neighbors, and requires use of the extended version of the
theory \cite{8} that describes the rounding of the ergodic to nonergodic
transition). Thus, the observability of the idealized MCT \cite{2} is
restricted to about one decade in $T/T_{c}-1$, showing the limitations of
this theory when applied to polymers \cite{15,18,19}.

In the $\alpha$-regime $F_{q}(t)$ can be fitted to the KWW law with an
exponent $\beta $ that is also weakly $q$-dependent ($\beta _{q}=0.70\pm 0.08$
for $q=6.9$ \cite{15}); for a detailed analysis of this $q$-dependence see
Ref. \cite{19}). The predictions of MCT for the $\beta $ regime can also be
tested in detail and compare rather favorably with MCT \cite{19}, although
the same caveat over the restricted ``temperature window'' where MCT is
applicable must again be made.

This analysis can be replaced for several pressures and thus one can trace
out a ``critical line'' $T_{c}(p)$ in the ($T,p$) plane separating the
liquid from the ideal glass (Fig. 4). Carrying out simulations at constant
pressure and at constant density that lead to the same point ($T_{c}(p),p)$
on the line, one finds that indeed $\tau _{q}\propto (T-T_{c})^{-\gamma _{q}}
$ holds with the same $T_{c}$ and the same $\gamma _{q}$ for both paths in
the ($T,p$) plane. This underscores that the MCT critical line is indeed
physically significant.

Also the bond fluctuation model has been compared to MCT, both in its
idealized \cite{20} and extended \cite{21} version. While a fit of various
relaxation times and of the selfdiffusion constant to the VF law is nicely
compatible with the data and yields $T_{\text{VF}}\approx 0.125\pm 0.005$ 
\cite{22}, the MCT fits both \cite{20,21} yield $T_{c}\approx 0.15$, and
power law behavior occurs over a similar temperature interval as for the
continuum model. One characteristic difference between both models concerns
the $\beta $-relaxation regime, however: while in the off-lattice model the
first decay of the structure factor $F_{q}(t)$ \{Fig. 2\} is due to
small-amplitude motions (over distances of the order of 10\% of the
distances between effective monomers and their nearest neighbors), no such
small scale motion is possible in the bond fluctuation model: either a
monomer hops a lattice unit), or it cannot hop at all. This distinction also
shows up when we compare the time dependence of the scaled mean square
displacements of monomers for both models (Fig. 5). Both the lattice model
and the continuum model are thought to correspond essentially to the same
physical system, a dense melt of short polymers, on a coarse-grained level.
This implies that on mesoscopic scales (length scales of the order of the
distance between effective monomers or larger) these different
coarse-grained models should yield very similar results. Comparing suitably
scaled data, so that units of length and time are absorbed in the scales,
this is indeed the case (Fig. 3). Note that the regime where MCT is
applicable is actually the regime of rather small displacements, up to the
order of distances between monomers,from where on then a Rouse-like behavior
controls the dynamics.

\section{Comments on other theoretical concepts and some concluding remarks}

A particular advantage of the lattice model is that the slithering snake
algorithm allows to obtain equilibrium properties of the model at fairly low
temperatures, and also the configurational entropy $S(T)$ can be computed 
\cite{23} in order to test the Gibbs-Di Marzio theory \cite{6}. Indeed one
finds that $S(T)$ decreases to about 1/3 of its high temperature value $S$($%
\infty )$ as $T$ approaches $T_{g}$, but there the curve $S(T)$ vs $T$ bends
over and gives evidence that $S(T)$ stays nonzero at lower temperatures.
Conversely, if one extracts the quantities that are used in the theory \cite
{6} directly from the simulation and inserts them into the (approximate!)
theoretical formula \cite{6,23}, one would find an ``entropy catastrophe'' \{%
$S(T)$ becoming negative\} at $T\lesssim 0.18$. This unphysical result,
however, is easily traced back to a severe underestimation of $S(\infty $)
by the approximations of \cite{6}. The ``entropy catastrophe'' \cite{6} thus
clearly is an artefact of an inaccurate approximation, one should not
attribute physical significance to it. In such manner simulations can go
beyond experiment for testing theories, because the input parameters of the
theories can be unambiguously extracted from the simulation as well, and a
comparison between theory and simulation is possible without adjustable
parameters whatsoever.

It is interesting to note that nevertheless the Adam-Gibbs equation \cite{7}%
, $D/D_{0}\propto \exp \{-E_{\text{act}}/TS(T)\}$, where $D_{0}$ is the
self-diffusion constant at very high temperatures and $E_{\text{act}}$ some
activation energy, provides a very good description of the diffusion
constants found in the simulation when one uses also the $S(T)$ found in the
simulation (Fig. 6). At the same time, the absence of any finite size effect
in Fig. 6 is at odds with the idea to attribute the slowing down implied by
the decrease of $D(T)$ to an increasing glass correlation length $\xi (T)$ 
\cite{9}, via $\tau (T)\propto [\xi (T)]^{z}$ with $z$ a dynamic exponent \cite
{10}: if such an hypothesis would hold, one would expect that decreasing L
should decrease $\xi (T)$ [since $\xi (T)\leq L$] and hence $\tau (T)$, and
in turn $D(T)$ should increase with increasing $L$. Fig 6 proves the absence
of such finite size effects. This finding is surprising, since a growing
static length can be extracted both from the pair correlation function in
the melt \cite{24} and from surface effects near hard walls \cite{25}. It
hence appears that there is a growing length identified in \cite{24,25} but
it is not responsible for the slowing down at $T_{g}$. Of course, there is
no contradiction with experimental results which find that $T_{g}$ in thin
films (or in pores, respectively) changes when the linear dimension of the
film (or pore respectively) is varied: depending on the boundary conditions
at the surface, the local mobility of monomers in the surface region
changes, and this effect is the more pronounced on the freezing the smaller
the linear dimension. No such surface effect is present with the periodic
boundary conditions in Fig. 6, of course. In experiments on the glass
transition in confined systems, finite size effects and surface effects can
never be clearly separated, unlike simulations where one can show that there
is no finite size effect (Fig. 6) at least in the temperature region studied
here, although there do exist surface effects \cite{25}.

Returning to MCT, we emphasize that idealized MCT does provide a good
description of a large number of simulation data, but only over a rather
restricted range of temperatures (about one decade in $T/T_{c}-1$) and
corresponding times (or viscosities, typically the range $10^{0}{}<\eta
(T)<10^{2}$ Poise if $\eta (T_{c})=10^{3}$ Poise: the huge range from 10$%
^{3}\leq \eta \leq 10^{13}$ Poise is outside the scope of the theory).
Correspondingly, only a small intermediate regime of small monomer
displacements in Fig. 5 is described - neither the initial increase that
depends on microscopic properties of the model, nor the regime of hopping
processes that lead to a Rouse-like relaxation of coil configurations
(before ultimately ordinary diffusion sets in) are part of the theory. Since
there occur smooth crossovers between the various regimes, a reliable
assessment of the validity of MCT for polymer melts near $T_{g}$ is
difficult. A more complete theory (that unifies e.g.\ MCT and the Rouse
model) would be very desirable.

\vspace{5mm}

\noindent
\underline{\bf Acknowledgements:} We are grateful to S. B\"{o}hmer, K. Okun, V.
Tries and M. Wolf\-gardt, who have contributed to earlier stages of this work,
for their fruitful collaboration. We acknowledge the financial support from
the Deutsche Forschungsgemeinschaft (DFG), grant No SFB262/D2, and we thank
the H\"{o}chstleistungsrechenzentrum J\"{u}lich (HLRZ) and the Regionales
Hochschulrechenzentrum Kaiserslautern (RHRK) for generous grants of computer
time.

\begin{figure}[h]
\caption{Structure factor $S(q)$ plotted vs wavenumber $q$, for a system of
120 off-lattice bead-spring chains with chain length $N=10$, simulated in
the NpT-ensemble at scaled pressure $p=1$, choosing Lennard-Jones units $%
\varepsilon =1,\sigma =1$ (and $k_{\text{B}}$=1). Note that the zero of the
ordinate for each curve is shifted upward by 0.001 relative to the previous
one. From Bennemann et al. \protect\cite{15}.}
\label{fig1}
\end{figure}

\begin{figure}[h]
\caption{Dynamic structure factor $F_{q}(t)$ of the off-lattice model at
scaled pressure $p=1$ plotted vs a rescaled time $t/\tau _{q}$ (where $\tau
_{q}$ is defined from $F_{q}(t=\tau _{q})=0.3$) for $q=6.9$, the peak
position of the static structure factor at low $T$ (cf.\ Fig. 1). Only
temperatures in the range $0.46 \leq T\leq 0.7$ are included, as indicated.
From Bennemann et al. \protect\cite{15}.}
\label{fig2}
\end{figure}

\begin{figure}[h]
\caption{Log-log plot of the relaxation time $\tau _{q}$ versus $T-T_{c}$
with $T_{c}=0.45$, for a pressure $p=1$ and three values of $q$. The
straight lines indicate the power laws $\tau _{q}\propto (T-T_{c})^{-\gamma
_{q}}$ with exponents $\gamma _{q}=2.3$, 2.1 and 2.0 (from above to below).
From Bennemann et al. \protect\cite{18}.}
\label{fig3}
\end{figure}

\begin{figure}[bh]
\caption{Critical line $T_{c}(p)$ in the ($T,p$) plane, for the off-lattice
bead spring model. From Bennemann et al. \protect\cite{18}.}
\label{fig4}
\end{figure}

\begin{figure}[h]
\caption{Comparison of the mean-square displacement $g_{1}(t)$ of inner
monomers \{$g_{1}(t)\equiv \left\langle \left[ \vec{r}_{i}(t)-\vec{r}%
_{i}(0)\right] ^{2}\right\rangle \}$ plotted vs scaled time \{$Dt/\langle
R_g^{2}\rangle$, $D$ being the selfidiffusion constant and 
$\langle R_g^{2}\rangle$
the mean square gyration radius of the chains\} for the off-lattice model
(curve marked MD) with the corresponding Monte Carlo results for the bond
fluctuation model at three temperatures. }
\label{fig5}
\end{figure}

\begin{figure}[h]
\caption{Selfdiffusion constant of the bond fluctuation model plotted vs
inverse temperature, for $L\times L\times L$ lattices with periodic boundary
conditions and several lattice sizes$L$. The solid and the dotted line are
fits to the Vogel-Fulcher and Adam-Gibbs equation, respectively. Data taken
from Binder et al. \protect\cite{24}.}
\label{fig6}
\end{figure}
%
%
\input{epsf}
\setcounter{figure}{0}

\begin{figure}[h]
\begin{center}
\begin{minipage}[t]{100mm}
\epsfysize=90mm
\epsffile{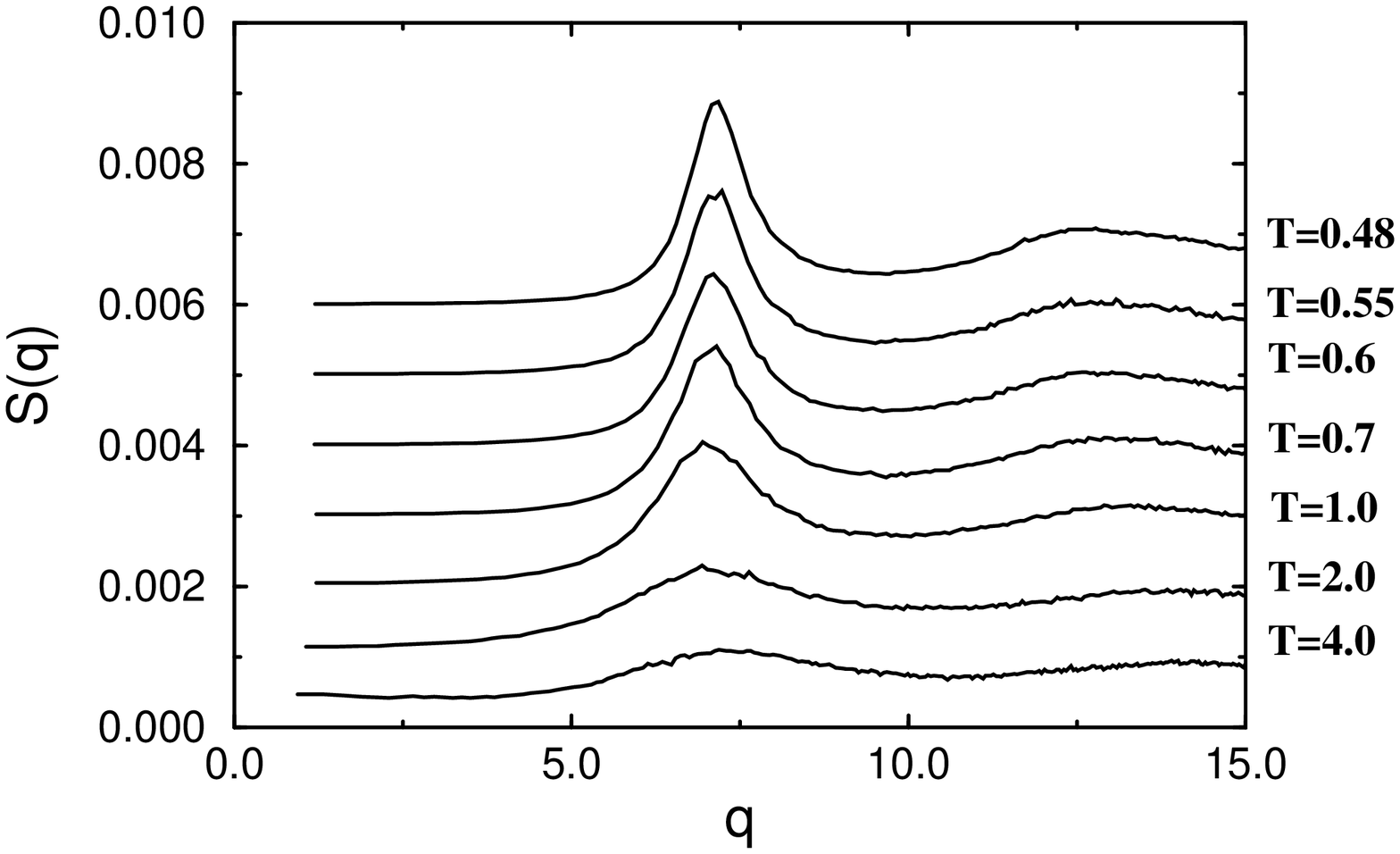}
\end{minipage}
\end{center}
\caption[]{}
\end{figure}

\begin{figure}[h]
\begin{center}
\begin{minipage}[t]{100mm}
\epsfysize=90mm
\epsffile{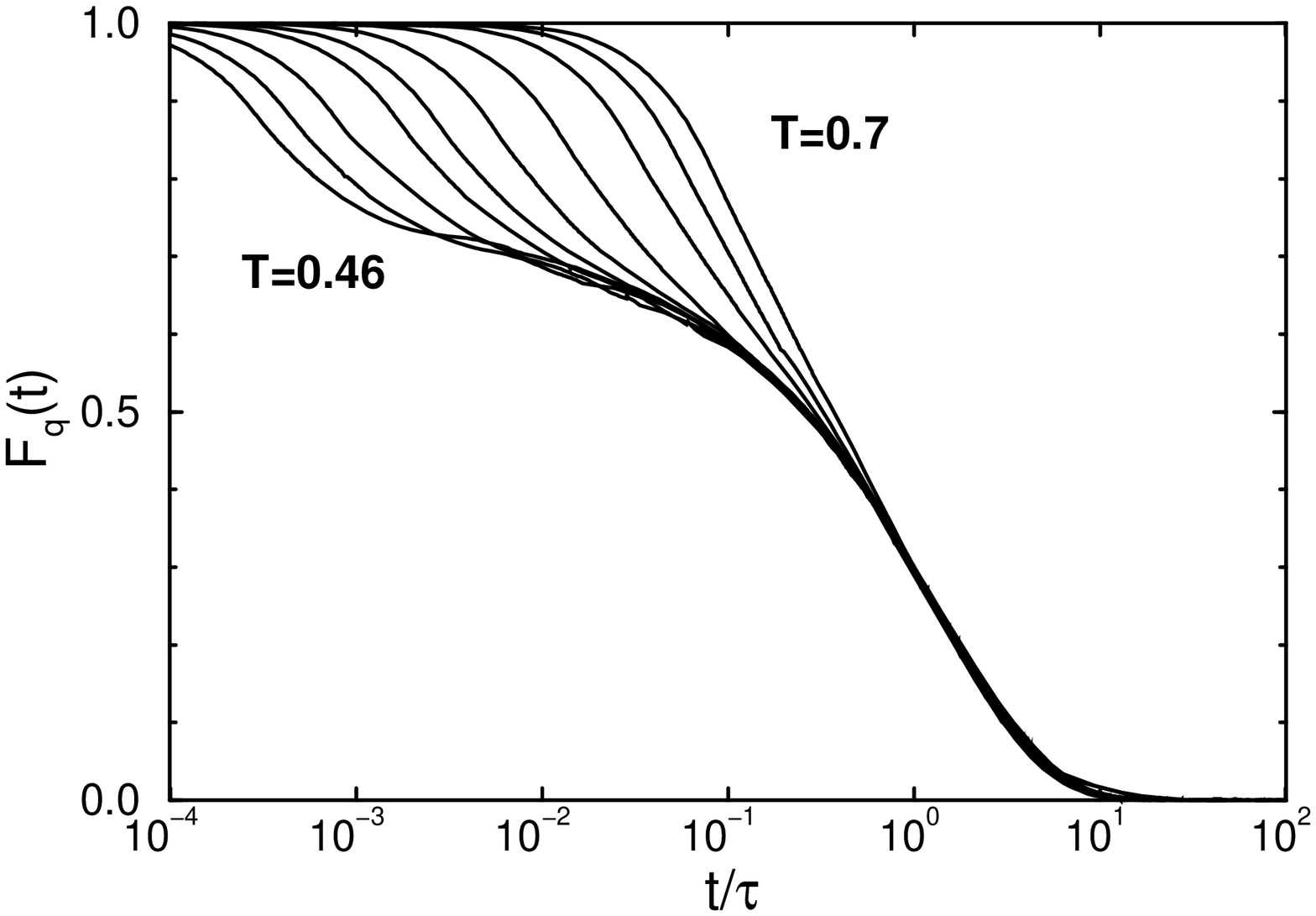}
\end{minipage}
\end{center}
\caption[]{}
\end{figure}

\begin{figure}[h]
\begin{center}
\begin{minipage}[t]{100mm}
\epsfysize=90mm
\epsffile{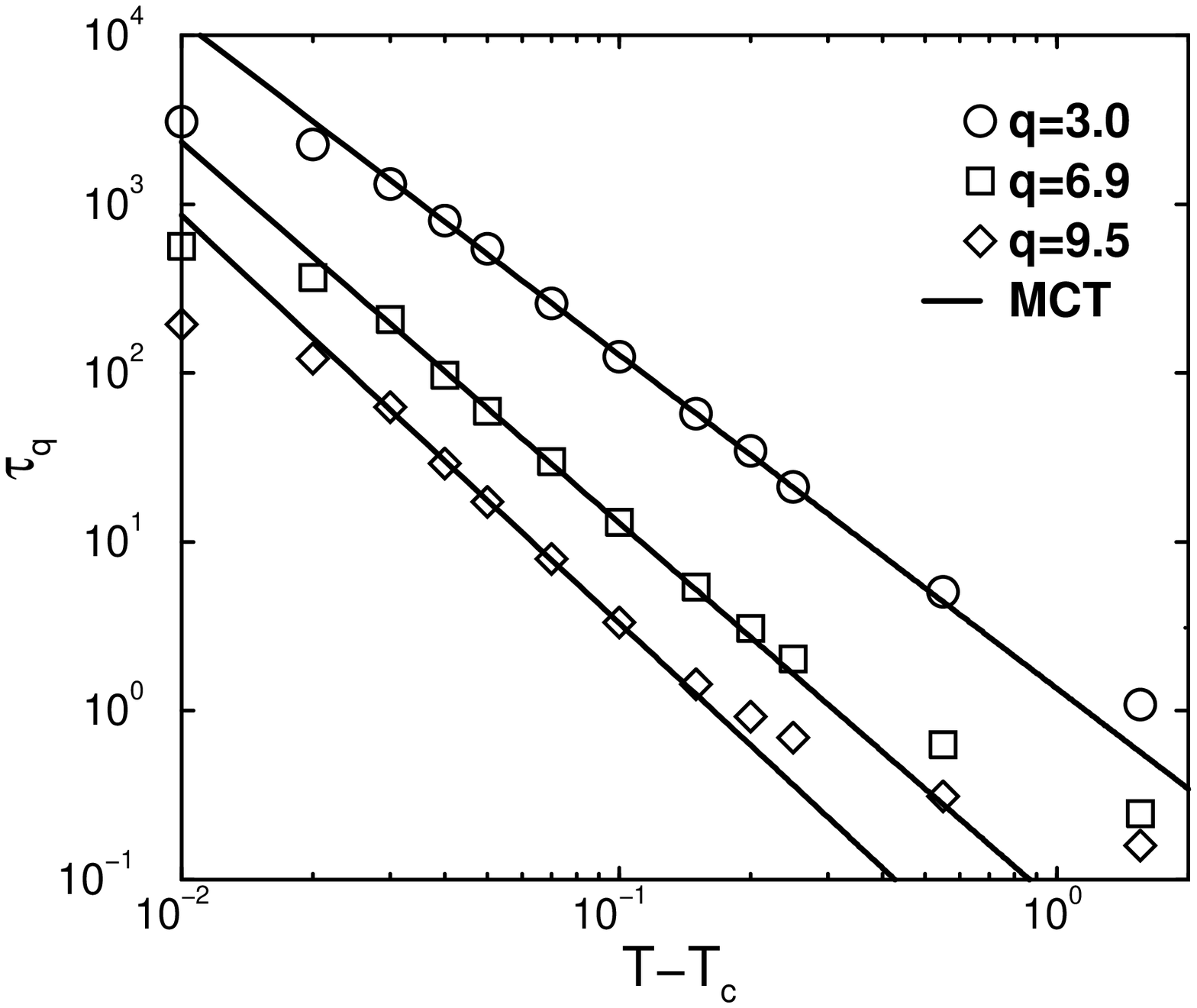}
\end{minipage}
\end{center}
\caption[]{}
\end{figure}

\begin{figure}[h]
\begin{center}
\begin{minipage}[t]{100mm}
\epsfysize=90mm
\epsffile{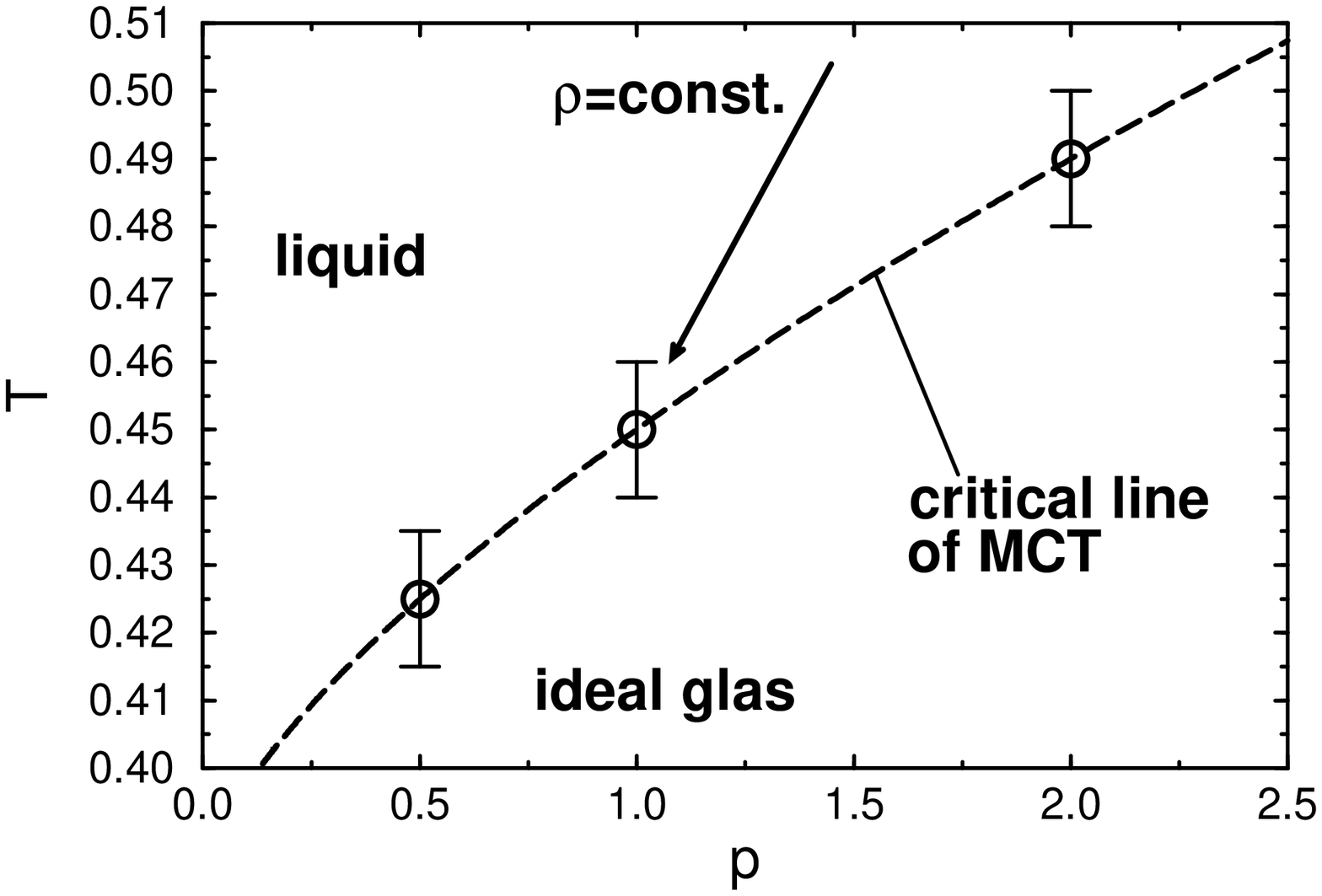}
\end{minipage}
\end{center}
\caption[]{}
\end{figure}

\begin{figure}[h]
\begin{center}
\begin{minipage}[t]{100mm}
\epsfysize=90mm
\epsffile{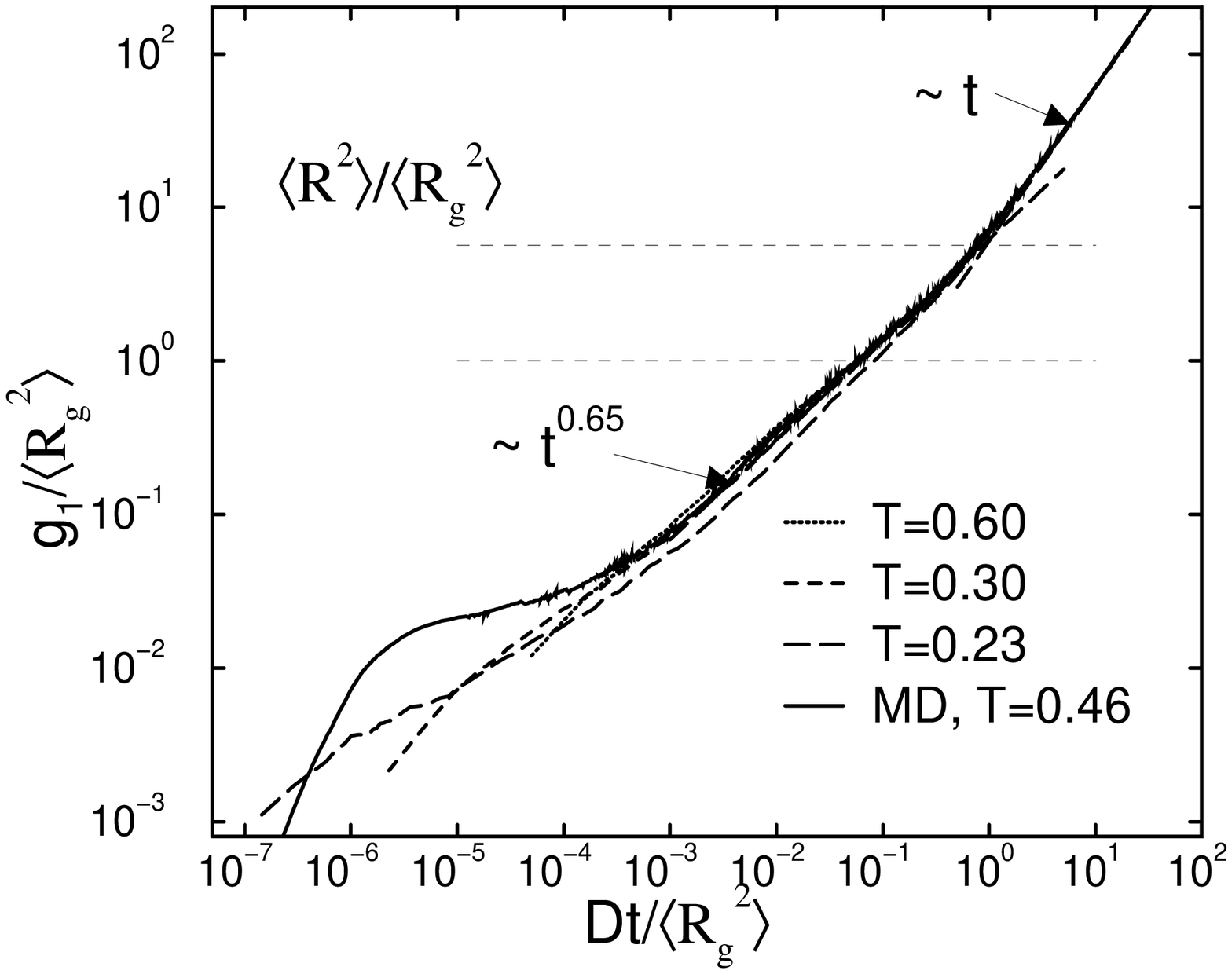}
\end{minipage}
\end{center}
\caption[]{}
\end{figure}

\begin{figure}[h]
\begin{center}
\begin{minipage}[t]{100mm}
\epsfysize=90mm
\epsffile{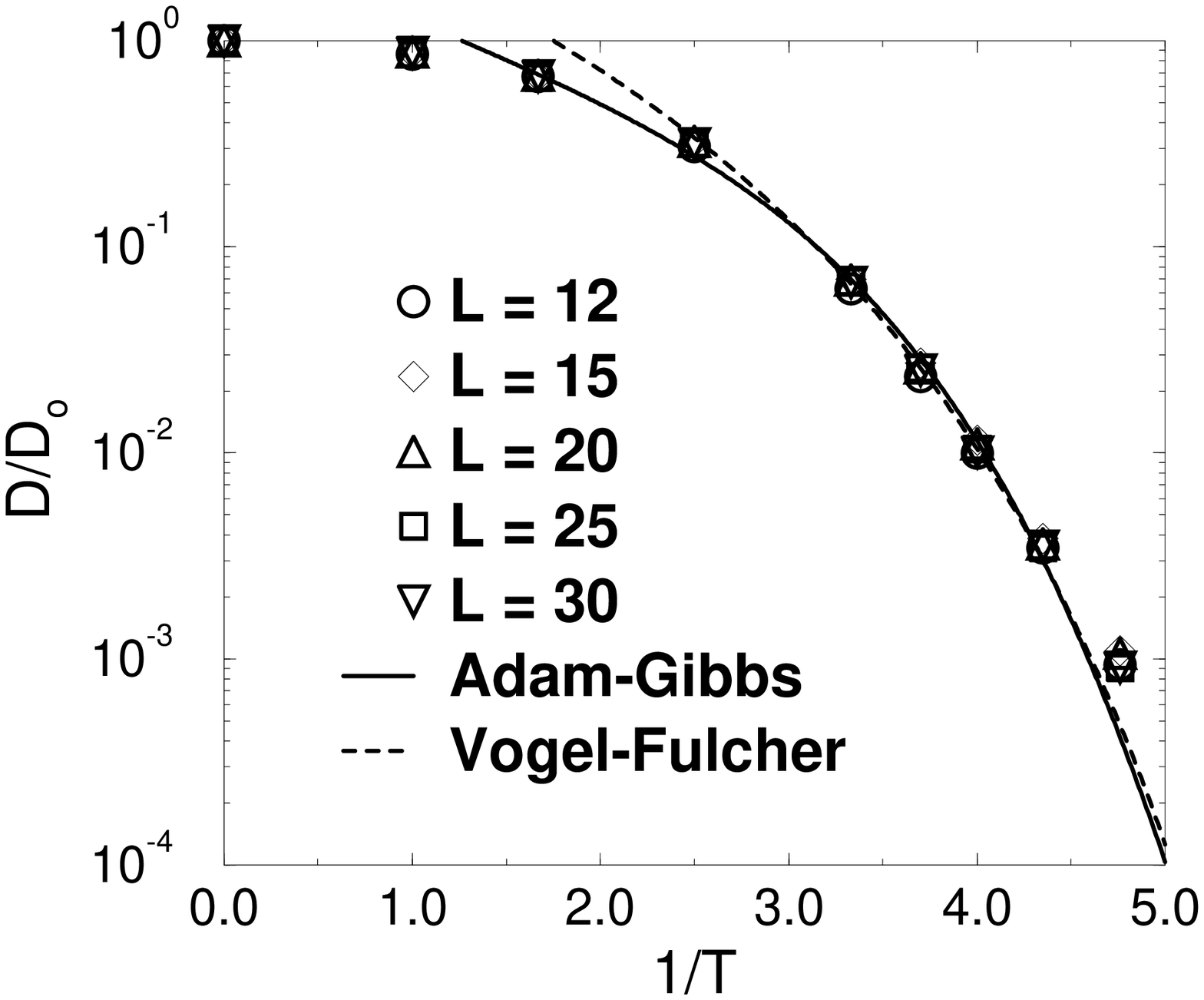}
\end{minipage}
\end{center}
\caption[]{}
\end{figure}

\end{document}